\documentstyle[aps,preprint]{revtex}
\begin{document}
\draft
\title {FISSION FRAGMENT SPECTROSCOPY ON A $^{28}$Si+$^{28}$Si
QUASI-MOLECULAR RESONANCE} 

\author { R. Nouicer, C. Beck, N. Aissaoui, T. Bellot, G. de France, D.
Disdier, G. Duch\`ene, \\
A. Elanique, R.M. Freeman, F. Haas, A. Hachem, \\ 
F. Hoellinger, D. Mahboub, D. Pr\'evost, V. Rauch } 

\address{\it Institut de Recherches Subatomiques, IN2P3-CNRS/Universit\'e Louis 
Pasteur,\\
B.P.28, F-67037 Strasbourg Cedex 2, France } 

\author { S.J. Sanders, T. Catterson, A. Dummer, F.W. Prosser }

\address {\it Department of Physics and Astronomy, University of Kansas,
Lawrence, USA }

\author { A. Szanto de Toledo }

\address {\it Depatamento de Fisica Nuclear, University of Sa\~o Paulo, Brazil }

\author { Sl. Cavallaro}

\address {\it Dipartimento di Fisica dell'Universit\'a di Catania, INFN
Catania, I-95129 Catania, Italy } 

\author { E. Uegaki}

\address {\it Department of Physics, Akita University, Akita 010, Japan}

\author { Y. Abe }

\address {\it Yukawa Institute for Theoretical Physics, Kyoto University,
Kyoto 606, Japan }

\date{\today}
\maketitle

\newpage

\begin{abstract}
{Fragment-fragment-$\gamma$ triple coincident measurements of the
$^{28}$Si~$+$~$^{28}$Si reaction at E$_{\rm lab.}$ = 111.6\ MeV carefully
chosen to populate J = 38$^{+}$ resonance have been performed at the VIVITRON
tandem facility by using Eurogam Phase II $\gamma$-ray spectrometer. In the
$^{28}$Si~$+$~$^{28}$Si exit-channel, the resonance behavior of the
$^{28}$Si~$+$~$^{28}$Si reaction at the beam energy is clearly confirmed. An
unexpected spin disalignment has been observed in the measured angular
distributions in the elastic, inelastic, and mutual excitation channels. This
disalignment is found to be consistent with particle-$\gamma$ angular
correlations and supported by the molecular model prediction of a ``butterfly
motion". The K$^{\pi}$ = 0$^{+}_{3}$ band corresponding to the large prolate
deformation of the $^{28}$Si is more intensely fed in the resonance region. The
selective population of high-excited states are discussed within a statistical
fusion-fission model. In the $^{32}$S $+$ $^{24}$Mg exit-channel, the
spectroscopic study of the $^{32}$S, has revealed the contribution of a new
$\gamma$-ray transition $0^{+} (8507.8\ keV) \to 2^{+}_{1} (2230.2\ keV) $.} 
\end{abstract}

\vskip 2.0cm
 
{ PACS number(s): 25.70.Ef, 25.70.Jj, 25.70.Gh, 25.70.Lm,24.60.Dr, 23.20.Lv }

\newpage
\centerline {\bf I. INTRODUCTION }
\vskip 1.1cm
The appearance of quasi-molecular resonances in light heavy-ion collisions such
as $^{28}$Si~$+$~$^{28}$Si [1] or $^{24}$Mg $+$ $^{24}$Mg [2] may open up the
possibility for studying nuclear structure at high spin in the continuum of the
combined complex. The large-angle yields in the elastic and inelastic
scattering channels show correlated narrow-width resonant structures, which
have been suggested to be associated with quasi-stable configurations with
extreme deformation [3]. This interpretation is supported by Nilsson-Strutinsky
model calculations of the potential energy surface, yielding high-spin prolate
shape isomers in the $^{56}$Ni-$^{48}$Cr mass region [4,5] and by the recent
discovery of a new region of the superdeformation for A~$\simeq$~60 [6]. 
\vskip -0.1cm
This very striking quasi-molecular resonant structure is possibly connected to
a rather unusual subset of high-spin states stabilized against the mixing into
the more numerous compound nucleus states by some special symmetry [3]. As a
matter of fact recent theoretical investigations have indicated that
shell-stabilized ``hyperdeformed" shapes may exist in the $^{56}$Ni nucleus
with large angular momenta [5]. By the use of a molecular model proposed by
Uegaki and Abe [7], a stable configuration of the dinuclear system is found to
be an equator-equator touching one, due to the oblate deformed shape of the
$^{28}$Si nuclei [8,9]. In this paper, we present for the first time
experimental results using very powerful coincidence techniques, which indicate
the possible occurence of a butterfly mode responsible for the quasi-molecular
resonant structure and the search for highly deformed bands in the $^{28}$Si
nucleus produced in the $^{28}$Si $+$ $^{28}$Si reaction at a resonance energy
E$_{\rm lab}$ = 111.6\ MeV. 
\newpage
\vglue -0.4cm
\centerline {\bf II. EXPERIMENTAL PROCEDURES }

\vskip 1.1 cm

The experiment was performed at the IReS Strasbourg VIVITRON tandem facility
with a $^{28}$Si beam at the bombarding energy E$_{\rm lab}$ ($^{28}$Si) =
111.6\ MeV carefully chosen to populate the well known 38$^{+}$ resonance [10].
The $^{28}$Si beam was used to bombard a 25 ${\rm \mu g /cm^{2}}$ thick
$^{28}$Si target. The thickness of the target corresponds to an energy loss of
130 keV smaller than the resonance width ($\Gamma_{c.m.} \approx $ 150\ keV).
The experiment has been carried out in triple coincidence modes
(fragment-fragment-$\gamma$). The fission fragments were detected in two pairs
of large-area position-sensitive Si(surface-barrier) detectors placed on either
side of the beam axis and their masses were determinated by using standard
kinematic coincidence techniques [11].The $\gamma$-rays were detected in
Eurogam Phase II multi-dectector array [12], which consists of 54
Compton-suppressed germanium (Ge) detectors, 30 tapered coaxial Ge detectors
from Eurogam Phase I located in the forward and backward hemispheres, and 24
clover detectors installed at $\sim $ 90$^{o}$ relative to the beam axis. The
number of the Ge crystals at each angle with respect to the beam direction is
(5, 22$^{o}$), (10, 46$^{o}$), (24, 71$^{o}$), (24, 80$^{o}$), (24, 100$^{o}$),
(24, 109$^{o}$) (10, 134$^{o}$) and (5, 158$^{o}$). Energy and relative
efficiency calibrations of Eurogam Phase II were obtained with standard
$\gamma$-ray sources and a AmBe source for the higher energy $\gamma$-ray
region [12,13]. 
\vglue 0.4cm

\newpage

\centerline {\bf III. EXPERIMENTAL RESULTS }
\vskip 1.1cm

In order to better understand the experimental results, we present in the
Fig.1 a two-dimensional energy spectrum E$_{3}$-E$_{4}$ of the
fragments in coincidence in the respective angular ranges 30.5$^{o} \le
\theta^{\rm lab.}_{3} \le $ 56.4$^{o}$ and 29.7$^{o} \le \theta^{\rm lab.}_{4}
\le $ 58.4$^{o}$. Three regions,noted 1, 2 and 3, can be easily distinguished
in the figure. The region 1 corresponds to the binary products of the $^{28}$Si
$+$ $^{28}$Si reaction. In this region the spectrum shows well structured
distributions indicating that the bombarding energy corresponds well to the
resonance energy. The region 2 arises from to the reaction products of the
$^{28}$Si with a contamination ($^{63}$Cu), and region 3 corresponds to the
light charged particles ($p$, $\alpha$) coincidences arising from the
fusion-evaporation of the $^{56}$Ni compound system. The selection of the
different exit channels has been obtained by using mass spectra constructed by
standard binary kinematic relations [12]. 
 
The $\gamma$-ray emitted by excited fragments are detected by the Ge detectors
of Eurogam Phase II. Doppler-shift corrections were applied to the $\gamma$-ray
data on an event-by-event basis using measured velocities of the detected 
fragments by using the following relativistic-like equation~:  
\begin{equation}
\small
E_{\gamma} = E_{0}
\left( 1 + \beta \ \cos \vartheta_{D} - {1 \over 2}\ \beta^{2} \right)
\end{equation}
where $\cos \vartheta_{D}$ is Doppler angle. Since the velocities of the
fission fragments ($\beta$~($^{28}Si)~=~{v \over c}~=$ 7.4~$\%$) are larger
than the velocities of the evaporation residues ($\beta = { v \over c} =$ 2
$\%$), then the equation~(1) is well adapted for fusion-fission (FF) reactions.

\vskip 0.5cm

\centerline{\large \bf A - Coincidence measurement for $^{28}$Si $+$ $^{28}$Si
exit channel } 

\vskip 0.3cm

Reaction $Q$-value spectra were obtained for the different binary-reaction
channels based on the known entrance-channel parameters, the deduced mass of
the fragments, and the measured fragment angles. In Fig.2 the
experimental excitation-energy (E$_{\rm x}$) spectrum for $^{28}$Si $+$
$^{28}$Si exit channel is displayed by points and the curves are for the
calculations of the transition state model (TSM) for fission decay. For E$_{\rm
X}$ values less than 12 MeV the identification of the peaks is clear.
Pronounced structures are still observed in this spectrum at high excitation
energies. Since the number of possible mutual excitations becomes large above
12 MeV, the TSM calculations are in good agreement with the data and suggest
the dominance of fission decay at high excitation energy. The disagreement for
the low-lying states is due to the resonant behavior of this exit-channel. 

The $^{28}$Si($^{28}$Si,$^{28}$Si)$^{28}$Si identical particle exit-channel at
E$_{\rm lab.}$ = 111.6 MeV was found to have at backward angles (between
70$^{o} \le \theta_{c.m.} \le $ 110$^{o}$) strongly oscillatory angular
distributions for the elastic, inelastic, and mutual excitation channels as
shown in Fig.3. The present large-angle high-quality data, with
good position-resolution and high statistics, are well described by the curves
of Fig.3 as calculated by [P$_{L}(\cos \theta_{c.m.} )$]$^{2}$
shapes with L = 38 ${\rm \hbar}$ in perfect agreement in the elastic channel
with the previous data of Betts et al. [10]. The fact that the measured angular
distributions in the elastic, inelastic, and mutual excitation channels
correspond to shapes characterized by the same single Legendre polynomial
squared means that the resonant behavior is dominated by a unique and pure
partial wave associated with the angular momentum value L = 38 ${\rm \hbar}$.
This value can finally be considered as the spin of the well defined and
isolated quasi-molecular resonance. The dominance of the angular momentum L =
38 ${\rm \hbar}$ in these three resonant channels implies that the projection
of the spin along the direction perpendicular to the reaction plane is {\it m
= 0}. 

In order to understand this disalignment we will focus our analysis on the
fragment-fragment-$\gamma$ coincidence data of the $^{28}$Si + $^{28}$Si exit
channel. Spin-alignment estimations of the low-lying excitation states (single
inelastic 2$^{+}_{1}$ and mutual inelastic (2$^{+}_{1}$, 2$^{+}_{1}$)
exit-channels) have been deduced by measuring their particle-$\gamma$ angular
correlations with Eurogam Phase II. Three quantization axes have been defined
as follows : a) corresponds to the beam axis, b) axis normal to the scattering
plane, and c) axis perpendicular to the a) and b) axes. The fragment detectors
are placed symmetrical with respect to the beam axis and their centers are
located at angles $\theta_{\rm lab.}$~$\simeq$~45$^{o}$, then the c) axis can be
approximatively considered as the molecular axis parallel to the relative
vector between the two centers of the out-going binary fragments. In
Fig.4 the results of the $\gamma$-ray angular correlations for the
mutual excitation exit-channel are shown. The minima observed in a) and b) at
90$^{0}$ imply that the intrinsic spin vectors of the 2$^{+}$ states lie in the
reaction plane and are perpendicular to the orbital angular momentum. So the
value of the angular momentum remains close to L = 38 ${\rm \hbar}$ for the two
exit channels, in agreement with Fig.3. The maximum around 90$^{0}$
in c) suggests that the $^{28}$Si spin vectors are parallel to the fragment
directions with opposite directions. Such disalignments of the fragment spins
are, of course, not usual in deep inelastic processes [19], but a very long
life-time of the resonance might allow large microscopic fluctuations.
Fig.5 displays theoretical analyses for the normal-mode motions by
the molecular model [8], where the panel B) is found to have an extreme
concentration of the probabilities into channel spin of I = 0. In
$^{28}$Si~$+$~$^{28}$Si (oblate-oblate), the system would be dominated by {\it
butterfly motion} [8,9] and the theoretical results support the {\it spin
disalignments}. This is in contrast with the observed alignment for the
prolate-prolate system $^{24}$Mg $+$ $^{24}$Mg~[14]. 

The feedings of the $^{28}$Si states are measured in two angular regions (see
Fig.3). The first is the resonance region (73.2$^{o} \le
\theta_{c.m.} \le$ 105.6$^{o}$ angular range) and the second is the direct-like
region (56.8$^{o}\le \theta_{c.m.} \le$ 67.6$^{o}$ angular range).
Fig.6 shows the comparison between the experimental feedings
(diamonds and triangles) for states in $^{28}$Si populated in the resonance
region and the theoretical feedings (histograms) predicted by TSM. We can
remark that TSM does not reproduce well the experimental data feeding of the
0$^{+}_{1}$(K$^{\pi}$ = 0$^{+}_{1}$), 2$^{+}_{1}$(K$^{\pi}$ = 0$^{+}_{1}$),
4$^{+}_{3}$(K$^{\pi}$ = 0$^{+}_{3}$) and 4$^{-}_{1}$(K$^{\pi}$ = 3$^{-}_{1}$)
states of $^{28}$Si. The disagreement between experimental data and model
calculations for the feeding of the 0$^{+}_{1}$ and 2$^{+}_{1}$ states of
$^{28}$Si ground band is expected to be due to resonant effects. But the
disagreement for the feeding of the 4$^{+}_{3}$(K$^{\pi}$~=~0$^{+}_{3}$) and
4$^{-}_{1}$(K$^{\pi}$ = 3$^{-}_{1}$) states is less evident to explain. Is that
linked to the resonant structure ? This might be true since these states do not
appear in the direct-like region. 

One of the more interesting results in the K$^{\pi}$ = 0$^{+}_{1}$ band is that
the mutually excited states are more fed that the singly excited states
(0$^{+}_{1}$, 0$^{+}_{1}$), (2$^{+}_{1}$, 0$^{+}_{1}$) and (4$^{+}_{1}$,
0$^{+}_{1}$) (see Fig.6 : triangles) and the relative ratio
(extracted by fragment-fragment-$\gamma$ measurement) between them are equal to
13~$\%$. The experimental comparison of the feeding of $^{28}$Si states between
the resonance and direct-like regions is illustrated by Fig.6 and
Fig.7. The feeding of K$^{\pi}$ = 3$^{-}_{1}$ band is stronger than
for the K$^{\pi}$ = 3$^{+}_{1}$ band. The strong feeding of the K$^{\pi}$ =
3$^{-}_{1}$ is approximatively the same in the resonance region and direct-like
region. Finally a quantitative comparison is summarized in table 1. Where the
relative total feeding of K$^{\pi}$ bands are defined as follows : 

\begin{equation}
{\mathcal{R}}(K^{\pi}) = {{\mathcal{A}}(K^{\pi}) \over
{\mathcal{A}}(K^{\pi} = 0^{+}_{1}) }
\end{equation}

where ${\mathcal{A}}(K^{\pi})$ is the total feeding of a $K^{\pi}$ bands of the 
$^{28}$Si.

The strong feeding of the K$^{\pi}$ = 3$^{-}_{1}$ and K$^{\pi}$
= 0$^{+}_{1}$ bands implies that $^{28}$Si is dominated by oblate deformation.
The more unexpected result of the comparison is that of the K$^{\pi}$ =
0$^{+}_{3}$ corresponding to the large prolate deformation [15] is more fed in
the resonance region than in the direct-like region by a factor 6. 
\vskip 0.5cm
\centerline{\large \bf B - Coincidence measurement for $^{32}$S $+$ $^{24}$Mg exit channel }
\vskip 0.3cm
Fig.8 shows the excitation energy spectrum measured for the
$^{32}$S $+$ $^{24}$Mg exit-channel. Pronounced structures, are observed at
high excitation-energy. Since the number of possible mutual excitations becomes
very large above~11 MeV in this channel, the good agreement between
experimental data (points) and  TSM calculations (curves) for excitation-energy
spectrum (see Fig.8) at low energy implies that the reaction does
not proceed via resonant effects in this exit-channel. This confirms the lack
of resonant structures in excitation functions for the $^{28}$Si $+$ $^{28}$Si
$\to$ $^{32}$S $+$ $^{24}$Mg channel. The disagreement between the experimental
data and the TSM calculations at high excitation-energy has been investigated
by looking at the $\gamma$-ray spectra of the $^{24}$Mg and $^{32}$S nuclei
with a gate on the excitation-energy (E$_{\rm X} > $ 10\ MeV). It is shown that
the $\gamma$ transitions corresponding to the $^{24}$Mg and $^{32}$S nuclei in
this region E$_{\rm X} >$~10~MeV can be formed by the mutual states in a very
complicated manner that the TSM model cannot take into account due to a lack of
knowledge about the experimental states at very high E$_{\rm X}$. 

\hspace*{0.3cm}The study of the $^{32}$S + $^{24}$Mg exit channel using
fragment-fragment coincidences has been supported by the
fragment-fragment-$\gamma$ measurement. The $\gamma$-rays associated to
$^{32}$S and $^{24}$Mg~have been established [12]. The $\gamma$-ray spectrum of
$^{24}$Mg shows more $\gamma$-ray transitions between positive parity states
and also the appearance of the high excitation-energy 6$^{+}_{1}$(8.11\ MeV)
and 6$^{+}_{2}$(9.53\ MeV) states. In summary, the desexcitation of the
$^{24}$Mg nucleus during its rotation seems to feed the first rotational bands
K$^{\pi}$ = 0$^{+}$ and K$^{\pi}$ = 2$^{+}$, which correspond to the prolate
deformation. 
\par
\vspace*{-0.1cm}
\hspace*{0.3cm}A more unexpected result for $^{32}$S + $^{24}$Mg exit-channel
arises from the spectroscopic study of the $^{32}$S nucleus (see
Fig.9). According to the branching ratios of the 1$^{-}_{3} \to $
2$^{+}_{1}$ arising to the $^{32}$S desexcitation, which have been given by
Rogers and al. [16], there is an indication from our data of a signature of a
new $\gamma$-ray transition, 
\vspace*{-0.1cm}
$$ 
\displaystyle{0^{+} (8507.8\ keV) \to 2^{+}_{1} (2230.2\ keV)} 
$$
\vspace*{-0.1cm}
\hspace*{0.3cm}We can remark from the inspection of the $\gamma$-ray spectrum
of $^{32}$S, as shown in the Fig.9, a strong feeding of the
negative parity states with lower spins. These states have been selectively
excited in the $\alpha$ transfer reaction such as $^{28}$Si($^{6}$Li,
d)$^{32}$S [17] and $^{28}$Si($^{16}$O, $^{12}$C)$^{32}$S [18]. 

\newpage

\centerline {\bf IV. SUMMARY AND CONCLUSIONS }

\vskip 1.1cm

A high-resolution study of fragment-fragment-$\gamma$ data collected with 
the Eurogam Phase II multi-detector array 
for the $^{28}$Si~$+$~$^{28}$Si reaction at bombarding energy E$_{\rm lab.}$ = 111.6\ MeV 
has allowed us to extract very attractive and new results.  
\par
In the $^{28}$Si $+$ $^{28}$Si exit-channel, the resonant behavior 
of the $^{28}$Si $+$ $^{28}$Si exit-channel is 
clearly observed by the present fragment-fragment coincidence data. 
The more unexpected result is the spin disalignment of 
the $^{28}$Si $+$ $^{28}$Si 
resonance. This has been demonstrated first by the measured angular distributions 
of the elastic, inelastic, and mutual excitation channels, which are  
dominated by a unique and pure partial wave with L = 38\ ${\rm \hbar}$, 
and has been confirmed by measuring their particle-$\gamma$ angular correlations 
with Eurogam Phase II. The disalignment 
is supported by molecular model predictions, 
in which the state with the butterfly mode is expected to correspond to the 
observed resonance.
The study of the feeding of bands of $^{28}$Si, has revealed that    
K$^{\pi}$ = 0$^{+}_{3}$ corresponding to the large prolate
deformation is more fed in the resonance region than in the direct-like region.
The good agreement between data and TSM (fusion-fission model) calculations suggests 
the importance of the fission decay at high-excitation energy while the resonant 
behavior involves the low-lying states.
\par
In the $^{32}$S $+$ $^{24}$Mg exit-channel, the spectroscopic study of the 
$^{32}$S nucleus, has revealed the contribution of a new $\gamma$ transition 
$0^{+} (8507.8\ keV) \to 2^{+}_{1} (2230.2\ keV) $ in the deexcitation process of  
$^{32}$S. The $\gamma$-ray spectrum of the $^{32}$S shows more $\gamma$-ray transitions 
between negative parity states with lower spins. 
\vskip 0.7cm
\noindent{\large \bf Acknowledgments} : We would like to acknowledge the 
VIVITRON operators for providing us well focussed $^{28}$Si beams. 
\newpage
\centerline{ \large \bf REFERENCES}
\vskip 1.1cm
\hspace*{-1cm} [1] R.R. Betts et al., Phys. Rev. Lett.  {\bf 47}, 23(1981).
\newline
\hspace*{-0.5cm} [2] R.W. Zurm\"{u}hle et al., Phys. Lett. {\bf B129}, 384(1983).
\newline
\hspace*{-0.5cm} [3] R.R. Betts et al., Nucl. Phys. {\bf A447}, 257(1985).
\newline
\hspace*{-0.5cm} [4] G. Leander et al., Nucl. Phys. {\bf A239}, 93(1975), 
M.E. Faber et al., Phys. Scr. {\bf 24}, 189(1981).
\newline
\hspace*{-0.5cm} [5] T. Bengtsson et al., Preprint Lund-MPh-{\bf 84/01} (1984).
\newline
\hspace*{-0.5cm} [6] C.E. Svensson et al., Phys. Rev. Lett.  {\bf 79}, 1233(1997). 
\newline
\hspace*{-0.5cm} [7] E. Uegaki et al., Phys. Lett. {\bf B231}, 28(1989) ; Prog. Theor. 
Phys. {\bf 90}, 615(1993).
\newline
\hspace*{-0.5cm} [8]  E. Uegaki et al., Phys. Lett. {\bf B340}, 143(1994).
\newline
\hspace*{-0.5cm} [9] E. Uegaki, Proc. of JAERI Symposium {\bf July 15-16}, Japan, (1997).
\newline
\hspace*{-0.7cm} [10] R.R. Betts et al., Phys. Lett. {\bf B100}, 177(1981).
\newline
\hspace*{-0.7cm} [11] R.R. Betts et al., Phys. Rev. Lett. {\bf 46}, 313(1981).
\newline
\hspace*{-0.7cm} [12] R. Nouicer, Ph.D. Thesis, Universit\'e Louis Pasteur, Strasbourg (unpublished).
\newline
\hspace*{-0.7cm} [13] C. Beck, R. Nouicer et al., Preprint IReS 97-22, {\bf nucl-ex/9708004} (1997)
\newline
\hspace*{-0.7cm} [14] A.H. Wuosmaa et al. Phys. Rev. Lett. {\bf 58}, 1312(1987).
\newline
\hspace*{-0.7cm} [15] F. Glatz et al., Zeit. f\"{u}r Phys. {\bf A303}, 239(1981).  
\newline
\hspace*{-0.7cm} [16] D.W.O. Rogers et al., Nucl. Phys. {\bf A281}, 345(1977)  
\newline
\hspace*{-0.7cm} [17] T. Tanabe et al., Phys. Rev. {\bf C24}, 2556(1981).  
\newline
\hspace*{-0.7cm} [18] C. Olmer et al., Proc. of Int. Conf. on Complex Nuclei,
Nashville (USA) {\bf V1}, 104(1974). 
\newline
\hspace*{-0.7cm} [19] J. Randrup, Nucl. Phys. {\bf A447}, 133(1985).

\vspace*{-0.5cm}
\begin{table}[hbpt]
\begin{center}
\hspace*{-1cm}\begin{tabular}{|c|c|c|c|c|c|c|} \hline
K$^{\pi}$ & 5$^{-}_{1}$ & 3$^{-}_{1}$ & 0$^{+}_{1}$ & 3$^{+}_{1}$ & 0$^{+}_{3}$ & 0$^{+}_{2}$\\
\hline
Kind of& large oblate&oblate&oblate&prolate&large prolate&prolate\\
deformation [15] &deformation&deformation&deformation&deformation&deformation &deformation\\
\hline
${\mathcal{R}}$(K$^{\pi}$)&{ } &{ }&{ }&{ }&{ }&{ } \\
resonance region&2.1 $\%$& 8.2 $\%$& 100 $\%$ & 4.4 $\%$ & 5.8 $\%$ & 1.2 $\%$ \\
\hline
${\mathcal{R}}$(K$^{\pi}$)&{ } &{ }&{ }&{ }&{ }&{ } \\
direct-like region &{0.7 $\%$ } &{6.7 $\%$ }&{ 100 $\%$ }&{ 4.5 $\%$ }&{ 1.1 $\%$ }&{1.3 $\%$ } \\
\hline
\end{tabular}

\vskip 0.7cm

\caption{\label{tab1} {\sl Relative total feeding of K$^{\pi}$ bands (${\mathcal{R}}$(K$^{\pi}$))
for states in $^{28}$Si populated in the resonance and in the direct-like regions. 
}}
\end{center}
\end{table}

\newpage

\centerline {\bf FIGURE CAPTIONS }

\vskip 1.1cm

Fig.1 : Two-dimensional energy spectrum E$_{4}$ versus E$_{3}$ of the
fragments detected by one pair of position-sensitive Si detectors for the
$^{28}$Si $+$ $^{28}$Si reaction at E$_{lab.}$~=~111.6~MeV. 

Fig.2 : Excitation-energy spectra for $^{28}$Si + $^{28}$Si exit-channel. The
points are the efficiency corrected experimental data. The curves are TSM
calculations for fission decay to particle-bound states (solid curve) and for
all decays (dotted curve). 

Fig.3 : Experimental angular distributions of the elastic, inelastic
2$^{+}_{1}$ and mutual excitation (2$^{+}_{1}$,~2$^{+}_{1}$), excitation
channels. The solid lines are calculated distributions based on a
$[$P$_{L}(\cos \theta_{c.m.} )]^{2}$ angular dependence. 

Fig.4 : Experimental $\gamma$-ray angular correlations of the
(2$^{+}_{1}$,~2$^{+}_{1}$) states in the resonance region 73.2$^{0}$ $\le
\theta_{c.m.}$ ($^{28}$Si) $\le$ 105.6$^{o}$ of the $^{28}$Si $+$ $^{28}$Si
exit channel for three quantization axes. 

Fig.5 : Molecular model prediction for Probability distributions of the
$^{28}$Si~$+$~$^{28}$Si ([ 2$^{+}_{1} \otimes $ 2$^{+}_{1}$] with L = J - I)
system versus channel spin I (see ref. [9]).

Fig.6 : Comparaison between experimental feeding (diamonds and triangles) for
states in $^{28}$Si populated in the resonance region and calculated feedings
predicted by TSM (histograms) for a statistical FF process.
 
Fig.7 : Experimental feeding (diamonds) for states in $^{28}$Si populated in
the direct-like region. 
 
Fig.8 : Excitation-energy spectra for $^{32}$S + $^{24}$Mg exit-channel. The
points are the efficiency corrected experimental data. The curves are TSM
calculations for fission decay to particle-bound states (solid curve) and for
all decays (dotted~curve). 
 
Fig.9 : $\gamma$-ray spectrum in coincidence with $^{32}$S for the $^{28}$Si
$+$ $^{28}$Si reaction at resonance energy E$_{lab}$ = 111.6\ MeV.

\end{document}